\documentclass[12pt,a4paper]{article}
\usepackage[compress,sort]{cite}
\usepackage{graphicx} 
\usepackage{amssymb}
\usepackage{amsmath}
\usepackage{fullpage}
\usepackage{hyperref}
\usepackage{microtype}

\begin{document}

\begin{titlepage}

\begin{flushright}
{MAN/HEP/2011/04 \\
MCnet-11-10}
\end{flushright}  
\vskip8mm

\begin{center}
{\Large \bf
Jet vetoing and Herwig++}
\vspace*{1.5cm}

Alex Schofield\footnote{\texttt{alex.schofield@hep.manchester.ac.uk}}\ ,
Michael H. Seymour\footnote{\texttt{michael.seymour@manchester.ac.uk}}
\\
School of Physics \& Astronomy, University of Manchester,\\
Oxford Road, Manchester, M13 9PL, U.K.

\bigskip
\bigskip

 { \bf Abstract }
\end{center}

\begin{quote}
We investigate the simulation of events with gaps between jets with a veto on additional radiation in the gap in Herwig++. We discover that the currently-used random treatment of radiation in the parton shower is generating some unphysical behaviour for wide-angle gluon emission in QCD $2\rightarrow2$ scatterings. We explore this behaviour quantitatively by making the same assumptions as the parton shower in the analytical calculation. We then modify the parton shower algorithm in order to correct the simulation of QCD radiation.
\end{quote}

\end{titlepage}

\newpage

\section{Introduction}
\label{sec:introduction}

	At modern colliders, such as the Large Hadron Collider and the Tevatron, a large fraction of the events produced are driven purely by the interactions of quantum chromodynamics (QCD). As a result of this, data generated in the early runs of the LHC is ideal for studying the nature of QCD. In addition to the availability of data, there have been many recent discoveries in the last ten years relating to the fundamental nature of QCD, such as non-global logarithms \cite{Dasgupta:2001sh,Appleby:2002ke} and super-leading logarithms \cite{Forshaw:2006fk,Forshaw:2008cq,Keates:2009dn}. From an experimental perspective, we also need a good understanding of QCD in order to reconstruct any physics of interest at said colliders. 

There are three fundamental theoretical problems associated with calculating observables for purely QCD processes. They are confinement, the large coupling constant for small energies, and the large logarithms generated by radiative corrections. These three properties make fixed order calculations unreliable, since each order can be as important numerically as that which preceded it. Therefore the alternate method of resummation, which in some approximation contains all orders of effects, needs to be used. The resummation of large logarithms can be done in two ways, either analytically or numerically. The analytical approach calculates the first order corrections in the soft limit and exponentiates to obtain the leading term at each order of the all orders result. The numerical approach, referred to as a parton shower, evolves the process from the hard scattering down to some infra-red cut-off scale by randomly generating radiation according to a physical principle. Formally the two approaches are equivalent in the large $N_c$ limit, where $N_c$ is the number of colours. There are a number of different parton shower programs currently available, see for example the recent review \cite{Buckley:2011ms}. In this paper we will focus on the event generator Herwig++ \cite{Gieseke:2011na,Bahr:2008pv}.

The gaps between jets process is an exclusive process which is ideal for working with early LHC data. By exclusive we mean that there is a restricted phase space for emissions of additional jets. In terms of the gaps between jets processes we impose a veto on radiation into the rapidity gap between the two hardest jets above a certain scale. From an experimental side there will be a large number of events produced since the underlying process is purely QCD $2\rightarrow2$ scattering. The signal is fairly simple since all we are looking for is a number of jets of at least two. While we are vetoing additional jets inside the gap there is still a possibility of emission into the region outside of the gap. From the theoretical side the basic process is well understood at the lowest order of perturbation theory, often referred to as the Born level, and approachable at all orders using a leading logarithmic resummation.  By varying the size of the rapidity gap and the amount of hadronic activity allowed it is possible to investigate a wide variety of QCD properties. In addition to this we can use this process to test how the parton shower, which is formally defined in the collinear limit, fares with regard to wide-angle radiation. Early LHC data on dijets with a veto have already been analysed by ATLAS with center of momentum energy $\sqrt{s}$=7$\,\mathrm{TeV}$ \cite{ATLAS-CONF-2010-085}.

The outline of the paper is as follows. First we summarise the current analytical approach to jet vetoing in QCD $2\rightarrow 2$ scatterings in section two, as described in Ref. \cite{Forshaw:2009fz}. Next, in section three, we explain how the analytical calculation differs from the parton shower picture in Herwig++. We then, in section four, explain how it is possible to modify the analytical calculation in order to quantitatively understand the parton shower behaviour. Moving on to section five, we use the insight gained from the analytical approach to modify the parton shower so that the unphysical behaviour is removed. Finally, we state and interpret our results in section six and then give a conclusion on the status of the parton shower in section seven. While our focus is on Herwig++, we mention in several places the similarities and differences with respect to other parton shower algorithms.

\section{Gaps between jets -- Analytical}
\label{sec:gbjanalytical}

\begin{figure}[!h] 
\begin{center}
\includegraphics[scale=1]{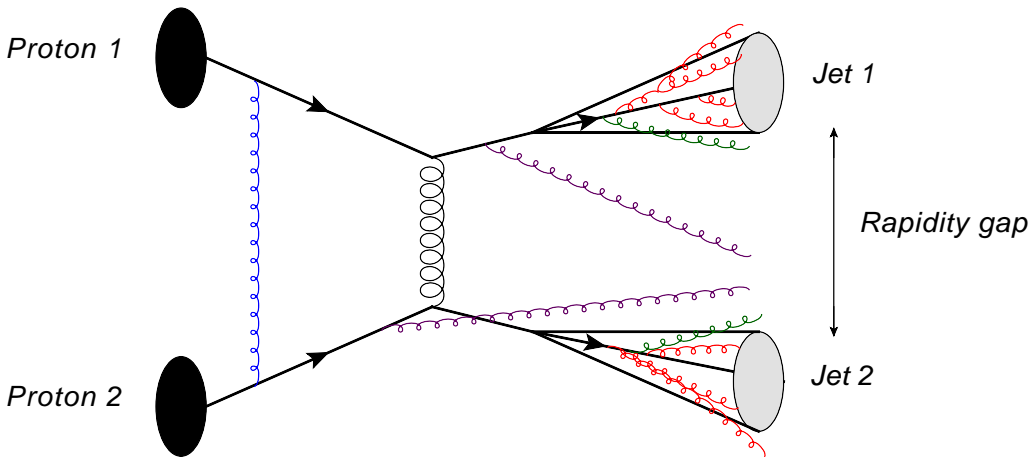}
\caption{Dijet production with a rapidity gap. The black lines represent the partonic interaction and those with arrows can be quarks, antiquarks or gluons. The coloured particles are generated by radiative corrections.}
\label{fig:process}
\end{center}	
\end{figure}

In this section, we summarise the work of Refs. \cite{Kidonakis:1998nf,Berger:2001ns,Appleby:2003sj,Forshaw:2009fz}. For the purpose of this paper we are mainly considering the colour structure, and as such will omit the details regarding the kinematics and convolution with PDFs. We will consider a partonic QCD $2\rightarrow 2$ scattering,

\begin{equation}
	p_{1} + p_{2} \rightarrow p_{3} + p_{4} ,
\end{equation}
where each $p_i$ is an arbitrary parton. We require that there are two high transverse momentum hard jets separated by a rapidity gap $Y$ with limited hadronic activity. An example $t$-channel process is shown in Figure \ref{fig:process}. We will refer to this as a gaps between jets event.

We restrict ourselves to the case of zero additional hard jets outside the gap. To quantify limited hadronic activity, we require that there are no jets in the gap above a scale $Q_0$, which we refer to as the veto scale. We will also define an additional scale, $Q$, which is the average transverse momentum of the two leading jets. The act of limiting the amount of radiation inside the gap induces a miscancellation between the real and virtual corrections which generates a large logarithm of the form

\begin{equation}
	L = \mathrm{Log}\left[\frac{Q}{Q_0}\right] .
\end{equation}
In this paper we will consider $Q$ to be some energy scale accessible by the LHC, and $Q_0$ to be small relative to $Q$ but large enough that perturbation theory can be used. Note that since we are considering radiation from the hard jets into the gap, we will only generate single soft logarithms. To perform the resummation, it will be useful to write the matrix elements in terms of a basis of colour states. The partonic cross section at the Born level can be written in the form

\begin{equation}
	\sigma_{0} = \mathrm{Tr[HS]} ,
\end{equation}	
where H is the hard matrix containing only Lorentz structure and S is the colour metric containing information about the interferences of the different colour flows. After the resummation of soft logarithms inside the gap\footnotemark\ we can write the partonic gap cross section as

\footnotetext{Here we are only considering the resummation of the logarithms which occur due to the miscancellation between the real and virtual emissions inside the gap. There are additional non-global logarithms\cite{Dasgupta:2001sh} resulting from secondary emissions into the gap which have not been resummed.}

\begin{equation}
	\sigma_{Gap} = \mathrm{Tr[H e^{-\xi\Gamma^{\dagger}} S e^{-\xi\Gamma}]} ,
\end{equation} 
where $\Gamma$ is the anomalous dimension which describes the evolution of the colour structure upon attaching an unresolved soft gluon and $\xi$ is given by

\begin{equation}
	\xi = \frac{2}{\pi} \int_{Q_0}^{Q} \frac{dk_t}{k_t} \alpha_s(k_t) . 
\end{equation}

The exponential factors serve as a Sudakov form factor \cite{Sudakov:1954sw}. If we know H, S and $\Gamma$ then we have determined the observable completely. The bulk of the work is that these objects are $M$-dimension matrices in colour space, where $M$ can be up to six for general SU($N_c$) theory, and if we wish to retain the full colour structure then each of the elements is a general function of~$N_c$.

In order to calculate these three objects we need to choose a specific basis for the colour states. For the analytical calculation the choice of an orthonormal basis is convenient as the colour metric becomes the unit matrix. This was the basis used by Ref. \cite{Forshaw:2009fz}, and is similar to the basis used in Ref. \cite{Kidonakis:1998nf}, but that basis was not normalized and did not benefit from the symmetric nature of the anomalous dimension in an orthonormal basis \cite{Seymour:2008xr}.  An alternate choice of basis, more suited to parton showers, which is the colour flow basis, will be discussed in the next section.

The hard matrix is determined by the Lorentz algebra of partial amplitudes \cite{Maltoni:2002mq}. Specifically, if we write the matrix element $M_i$ of the Feynman diagram $i$ as an expansion on a basis of colour states $\mid C_{j} \rangle$,

\begin{equation}
	M_i = \sum_{j} A^{(i)}_{j} \mid C_{j} \rangle ,
\end{equation}
then the hard matrix is

\begin{equation}
	\mathrm{H}_{ij} = \sum_{k,l} A^{(k)\dagger}_{i} A^{(l)}_{j},
\end{equation}
where the sums run over all possible Feynman diagrams contributing at the Born level. The soft metric can be determined by the projection,

\begin{equation}
	\mathrm{S}_{ij} = \langle C_{i} \mid C_{j} \rangle .
\end{equation}
By decomposing the colour basis states into Feynman diagrams, the components of the metric can be calculated by colour trace methods. The final task is to calculate the anomalous dimension, which can be written in colour basis independent notation as \cite{Forshaw:2008cq},

\begin{equation}
	\mathrm{\Gamma = \tfrac{1}{2}} Y  \mathrm{t}_t^2 + i \mathrm{\pi t_1 \cdot t_2 + \tfrac{1}{4} \rho (t_3^2 + t_4^2)} ,
\end{equation}
where $\mathrm{t}_i$ is a colour charge operator which quantifies the action of attaching a gluon to the parton $i$ in colour space. The form of $\mathrm{t}_i$ depends on the parton which is connected to the gluon and whether it is initial or final state. The first term, which is proportional to the size of the rapidity gap $Y$ and the colour exchange across the gap $\mathrm{t}_t^2$, is due to the emission of wide-angle gluons. This term serves as the main suppression factor for the resummed cross section. The second term is due to Coulomb gluons which appear in the virtual corrections that connect partons that are either both in the initial state or both in the final state. This term is mainly responsible for mixing between different colour basis states. The last term is an edge effect where small-angle radiation at the edge of a jet is able to make it into the gap. This term provides an additional small suppression. For large values of $Y$ ($Y>3$) and for a jet radius of $R=0.4$ then $\rho\sim0.6$ and independent of $Y$. In Figure \ref{fig:process} the wide-angle gluons are purple, the Coulomb gluons are blue and the gluons which generate the $\rho$ factor are green.

The calculation of the anomalous dimension can be done in a similar way to the colour metric. The anomalous dimension in an arbitrary colour basis is

\begin{equation}
	\Gamma_{ij} = \sum_{k} S^{-1}_{ik} \langle C_{k} \mid \Gamma \mid C_{j} \rangle .
\end{equation}

To compare to the experiment data, the resummed cross section needs to be convoluted with the PDFs. For a convenient choice of variables it is possible to factorise out the PDFs from the hard scattering and we are able to write the resummed hadronic cross section in the form,

\begin{equation}
	\sigma_{H}(Q,Q_0,Y) = \sum_{i} \mathcal{L}_{i}(Q,Y) \sigma_{Gap,i}(Q,Q_0,Y) ,
\end{equation}
where the summation is over all possible partonic processes and the function $\mathcal{L}_{i}$ contains the PDF integrals. In the eikonal limit the emission of gluons does not change the kinematics and therefore the PDFs are independent of $Q_0$ even after the resummation.

\section{Gaps between jets -- Parton Shower}
\label{sec:gbjps}
	
The Herwig++ parton shower, hereafter referred to as just the parton shower, is a sequence of quasi-collinear $1\rightarrow2$ splittings governed by the DGLAP equation \cite{Gribov:1972ri,Altarelli:1977zs,Dokshitzer:1977sg}. The shower works by evolving a scale $\tilde{q}^2$ \cite{Gieseke:2003rz}, related to the angle of the splitting. The DGLAP equations are solved using a random number generator in order to generate the scale of the next splitting, and the fraction of momenta to pass onto the two child partons. Here we will only consider the evolution scale and the colour structure of the parton shower.

The colour structure of the parton shower differs from the analytical calculation in a number of ways.  Before the parton shower begins, a hard process is generated in Herwig++ and a colour structure is chosen and retained throughout the whole evolution. The colour structures implemented in Herwig++ are a subset of the colour flow basis, where each basis state is a unique colour flow. In this basis, the colour metric is not orthonormal and thus there are interferences between colour states even at the Born level. To remove these interferences, the hard scattering is redefined in such a way that the Born level cross section is still generated \cite{Odagiri:1998ep}, but with the colour metric now being the unit matrix. Since the parton shower is probabilistic, the anomalous dimension is also taken to be real and diagonal. Thus we omit terms which are due to Coulomb gluons. This is equivalent to taking the large $N_c$ limit in the subsequent evolution\footnotemark, since interferences generated by Coulomb gluons will in general be suppressed by factors of $(1/N_c^2)$ relative to the leading contributions.\footnotetext{Coulomb gluons are not the only source of sub-leading colour interferences. There are also interferences between colour states suppressed by rapidity that exist even when Coulomb gluons are removed. However we have found that the difference between removing Coulomb gluons and taking the large $N_c$ limit is minimal.} Since the singlet colour structures have no associated initial hard scattering element, they will never contribute to physical processes in the large $N_c$ limit. Hence these singlet colour structures are not implemented in the parton shower. Note that though the parton shower works mainly in the large $N_c$ limit, there are also sub-leading corrections implemented in the colour charges of gluons and quarks, in order to generate the correct amount of radiation in the combined soft and collinear limit.

Most of the preceding discussion applies to all other general purpose event generators. Hard processes are generated exactly and then projected probabilitistically onto the colour flow basis (or another that is equivalent to it in the large $N_c$ limit) for subsequent evolution. However, in the next step, the evolution itself, one may draw a distinction between algorithms based on evolution of individual partons ("parton showers" like HERWIG, Herwig++ and the original virtuality-ordered shower in Pythia6\cite{Sjostrand:2006za}) or individual colour lines ("dipole showers", like the $p_t$-ordered shower in Pythia6 and Pythia8\cite{Sjostrand:2007gs}, Sherpa\cite{Gleisberg:2008ta}, Matchbox\cite{Platzer:2011bc}, Vincia\cite{Giele:2007di} and most other recent implementations).

\begin{figure}[!h] 
\begin{center}
\includegraphics[scale=0.8]{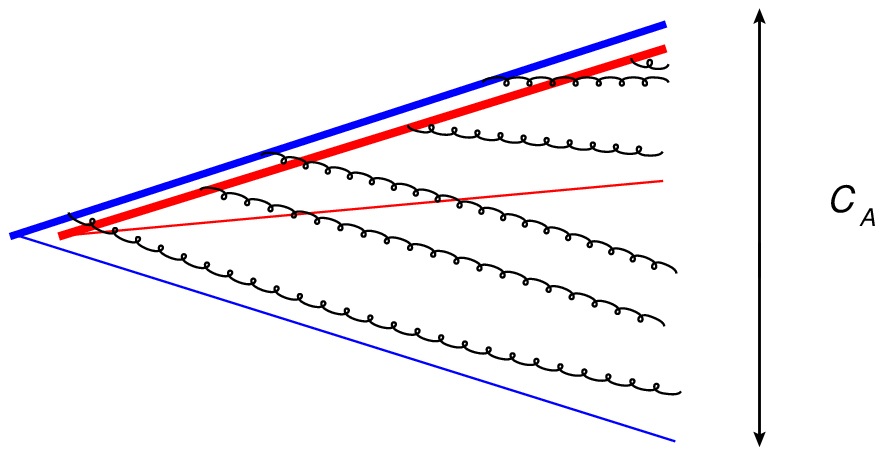}
\includegraphics[scale=0.8]{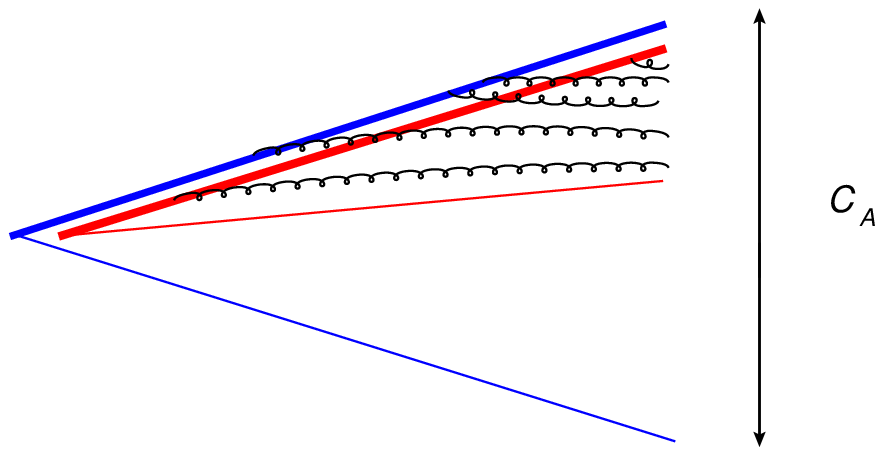}
\caption{Left: Wide-angle radiation is produced with a colour factor $C_A$ when the partner with the largest scale is chosen. Right: Only small-angle radiation is produced when the smaller scale is chosen. The blue and red lines are separated by wide and small angles respectively. The bold lines are the colour lines of the gluon and the faint lines are the colour lines of the partners.}
\label{fig:patterns}
\end{center}	
\end{figure}

In the large $N_c$ limit we can decompose the colour structure of a gluon into that of a quark line and an antiquark line. Any gluon produced in either the hard process or the subsequent shower will be connected by colour lines to two other partons. The scale of the radiation is proportional to the angle between the parton and the colour connected partner. For quarks there is only one possible partner, but for gluons there are two possible choices of partner. If we consider the case where one partner has a colour line scattered by almost $\pi$ and the other by a small angle, as occurs in, for example $qg\rightarrow qg$, then there is a large difference in the two possible initial evolution scales. An illustration of the two different radiation patterns is shown in Figure \ref{fig:patterns}. The current implementation in all three parton shower algorithms just mentioned is that the partner of a gluon is chosen from the two possibilities with 50-50 chance.  This implementation generates the correct amount of radiation for inclusive events. The radiation in exclusive events is, however, not treated correctly. For jet vetoing processes the amount of radiation into the gap is highly dependent on which of the two partners is chosen. 

\begin{figure}[!h] 
\begin{center}
\includegraphics[scale=1]{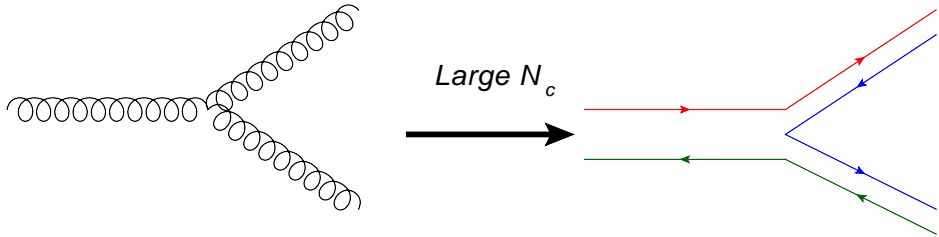}
\caption{Colour line representation of $g\rightarrow gg$ splitting in the large $N_c$ limit.}
\label{fig:splittingNc}
\end{center}	
\end{figure}

Once the partner is chosen the initial scale of radiation is determined and the parton shower begins to generate radiation. Each time a splitting occurs the colour line connections must be generated. For $q\rightarrow qg$ splitting there is only one possible colour connection. The parton shower picture for $g\rightarrow gg$ splitting is given in Figure \ref{fig:splittingNc}. There are two possible colour line connections, which can be thought of as either the colour line or anti-colour line splitting individually.  The parton shower chooses which line to split with a 50-50 choice. At the level of the shower this choice does not have any physical consequence. It is only when non-perturbative hadronization occurs that attaching radiation randomly can result in a different distribution of hadrons in the final state. By contrast, dipole showers evolve the two colour lines independently and the 50-50 choices of colour partner and emissions colour structure are not needed.

In addition to the above changes there are also additional effects implemented in the parton shower that are not present in the analytical calculation. Some non-global logarithmic effects are included in the parton shower as a result of the angular ordering \cite{Banfi:2006gy}. Energy-momentum conservation is absent from the analytical calculation due to the use of the eikonal approximation, but it is included in the parton shower. The implementation of energy-momentum conservation into the analytical calculation is discussed at the end of the next section.

\section{Modifications to the analytical calculation}
\label{sec:analyticalmodification}

To understand quantitatively how the parton shower currently behaves we can construct an analytical model of the colour evolution. This amounts to modifying our current analytical calculation in a number of ways. First we define a new hard scattering matrix $\mathrm{\overline{H}}$ such that its trace is equal to the Born cross section \cite{Odagiri:1998ep}, in the following way

\begin{equation}
	\mathrm{Tr[HS] \rightarrow Tr[\overline{H}]} ,
\end{equation}

\begin{equation}
	\mathrm{\overline{H}}_{ii} = \mathrm{H}_{ii} \mathrm{ Tr[HS] / Tr[H]},
\end{equation}
where we have normalized the colour basis states such that the diagonal elements of S are unity. We then take the large $N_c$ limit in the anomalous dimension and drop any imaginary terms. Since the anomalous dimension is now diagonal it can be treated as a rescaling factor for the hard structure of each of the independent colour flows. We can write the diagonal anomalous dimension in the form\footnotemark

\footnotetext{Here we only consider the colour structure and absorb the logarithmic factor into the definition of~$\Gamma$.}

\begin{equation}
	\Gamma = \Gamma_\rho + \sum_q \Gamma_q + \sum_g \Gamma_g  \leftrightarrow 
	e^{-\Gamma} = e^{-\Gamma_\rho} e^{-\sum_q \Gamma_q} e^{-\sum_g \Gamma_g},
\end{equation}
where we have factored out the wide-angle radiation from the small-angle radiation, and further factored out the contributions to the wide-angle radiation from the different types of parton and the sums over $i$ contain the contributions of all the partons of type $i$. For four external partons we will always have an even number of quarks and gluons separately and so we can treat the colour evolution of each parton type together. For the contributions to the anomalous dimension we only need to consider the number of colour lines crossing the gap. If a parton's colour line crosses the gap then it contributes a factor

\begin{equation}
	\Gamma_i = \tfrac{1}{4} C_i,
\end{equation}
where $C_i$ is $C_F$ for a quark line and $\tfrac{1}{2} C_A$ for a gluon line.  We implement the random choice of the partner in the analytical calculation by modifying the colour evolution factor. Since the parton shower generates colour evolution at the cross section level rather than at the amplitude level, and takes the anomalous dimensions to be diagonal, then we can simply double the anomalous dimension and take this to be our total colour evolution factor. At leading order in the anomalous dimension we can write the colour evolution factor for a single gluon as

\begin{equation}
\label{eq:halfmax} 
	e^{-2\Gamma_g} \simeq \tfrac{1}{2} (1 + e^{-4\Gamma_g}) ,
\end{equation}
which is the mathematical equivalent to the statement that half the time we have no wide-angle radiation and half the time we have twice as much. In the case where both partners are on the same side of the gap then the left-hand side of equation (\ref{eq:halfmax}) is used, but when we have one partner on each side of the gap then the right-hand side of equation (\ref{eq:halfmax}) is used.

From equation (\ref{eq:halfmax}) we can see that processes with hard gluons will tend to have a larger number of gap events than expected. For $gg\rightarrow gg$ there is a possibility that the choice of colour will be a quasi-singlet, where all four choose partners on the same side of the gap. We refer to this as a quasi-singlet since it has non-zero Born level hard components, whereas a physical singlet can only gain a hard component by radiative corrections interfering two colour flows. The quasi-singlet arises from the treatment of radiation in the parton shower, not because of the group properties of the theory. For large values of $L$ or $Y$, the suppressive factor then becomes

\begin{equation}
	e^{-2\Gamma_\rho} (\tfrac{1}{2} + \tfrac{1}{2}e^{-4\Gamma_g} )^4 \rightarrow \frac{1}{16} e^{- \xi C_A \rho} \sim  
	\frac{1}{16} \left(\frac{Q}{Q_0}\right)^{-\tfrac{2\alpha_s C_A \rho}{\pi}},
\end{equation}
which scales like a power law, where the last approximate equality holds only for a fixed coupling $\alpha_s$. In the analytical calculation the suppressive factor always has the form

\begin{equation}
	e^{-2\Gamma_\rho} e^{-8\Gamma_g} \sim \sum_{i} \left(\frac{Q}{Q_0}\right)^{-\tfrac{2\alpha_s C_A (\rho + A_i Y)}{\pi}} ,
\end{equation}
where $A_i$ are constants greater than zero and $i$ runs over all different colour flows. For large rapidity gaps this expression vanishes more quickly than that of the quasi-singlet. This quasi-singlet therefore results in a greater gap cross section at the more extreme edges of phase space. 

To illustrate the effects of these unphysical singlet type terms, we will look at the magnitudes of the different partonic processes in the resummation, with and without the modifications to the colour evolution. We define the resummed partonic fraction $\omega_i$ of a process $i$ to be,

\begin{equation}
	\omega_i = \frac{\sigma_{i,r}}{\sum_{j} \sigma_{j,r}},
\end{equation} 
where $\sigma_{i,r}$ is the resummed cross section for a given partonic process $i$.
\begin{figure}[!h] 
\begin{center}
\includegraphics[scale=0.5]{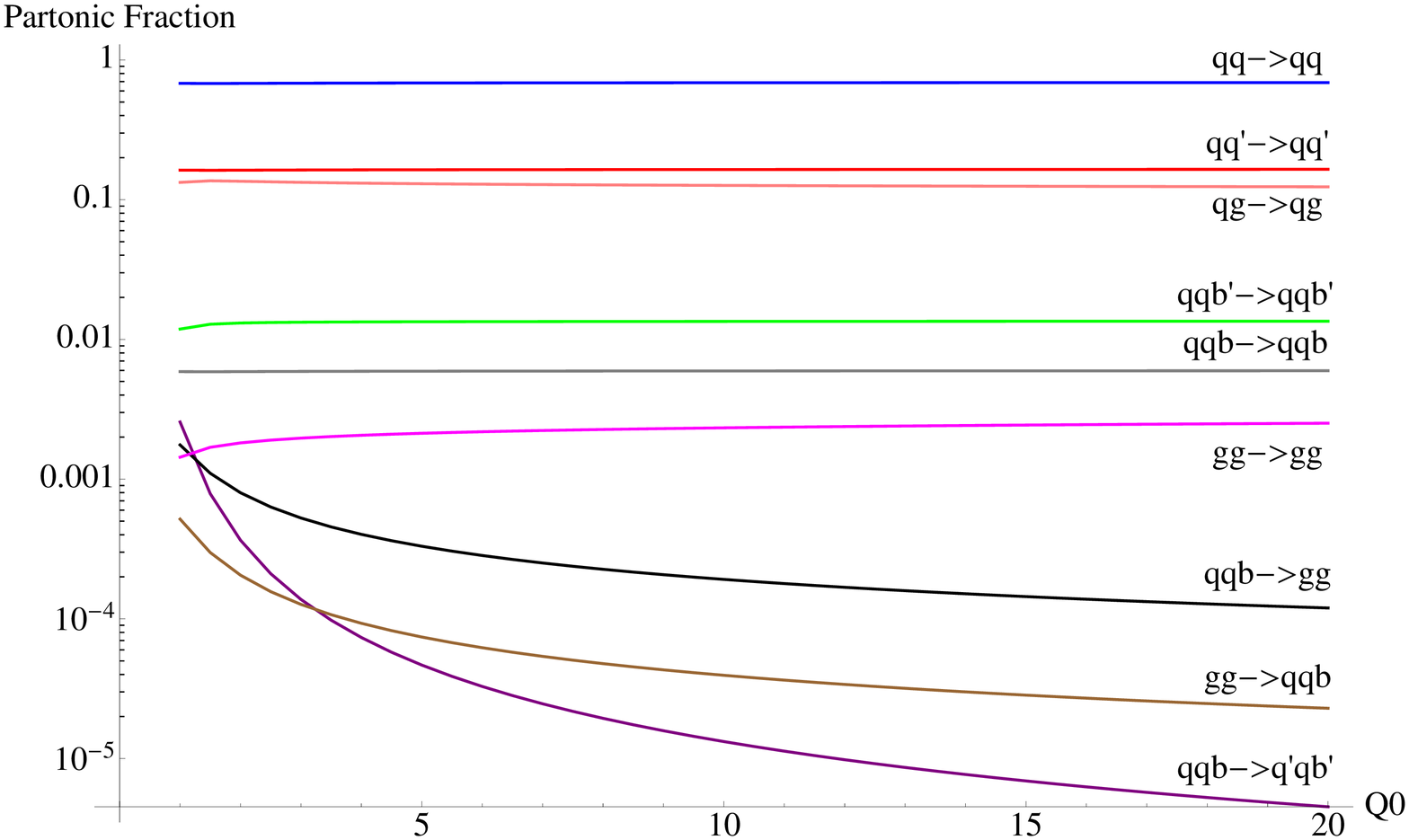}
\includegraphics[scale=0.5]{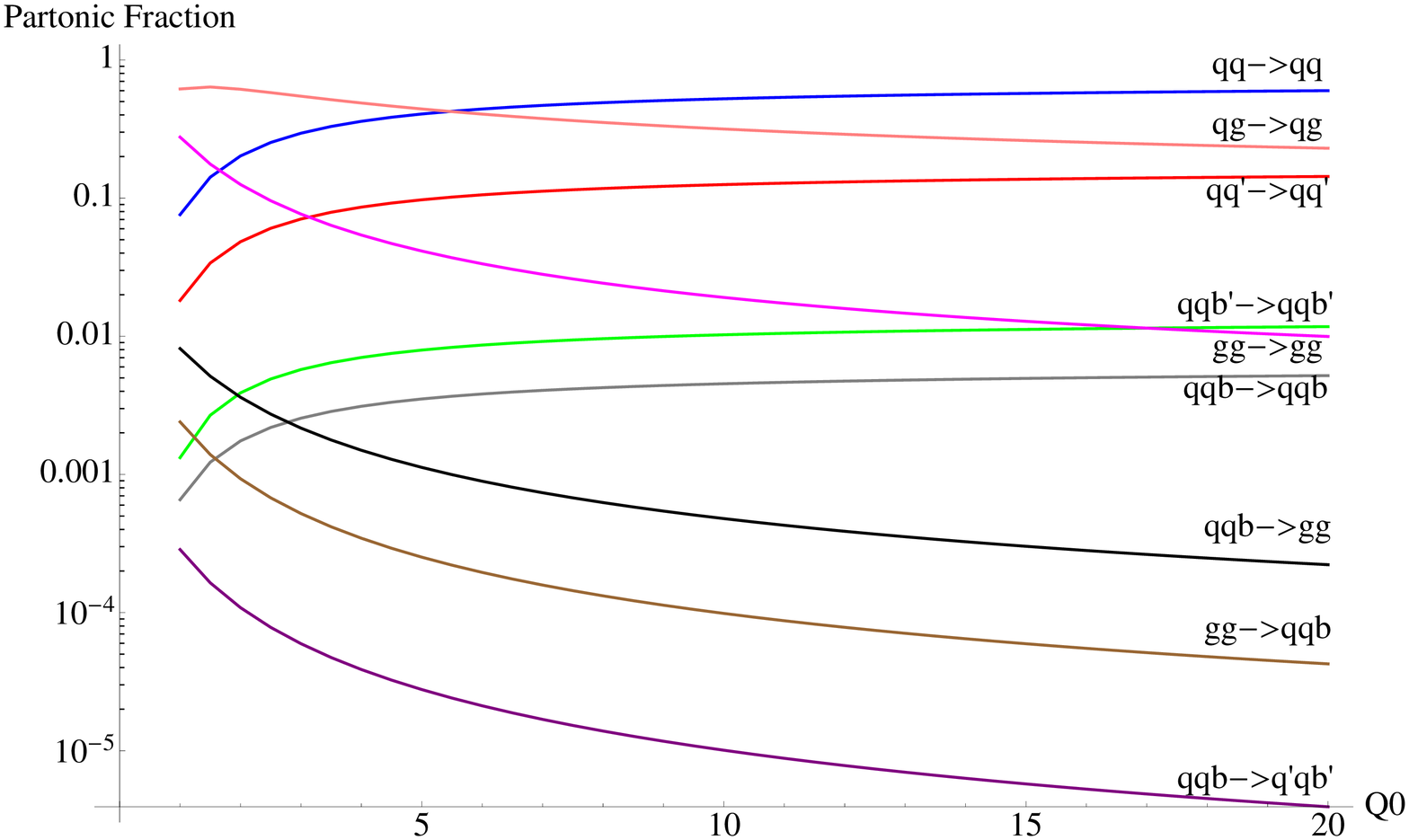}
\caption{Partonic fractions of the resummed cross section as a function of veto scale for $Y$=5, $Q$=500$\,\mathrm{GeV}$. The upper and lower plots are without and with the change to colour evolution respectively. The PDF sets used are MSTW08lo \cite{Martin:2009iq}. }
\label{fig:fractionplot}
\end{center}	
\end{figure}

The upper plot in figure \ref{fig:fractionplot} shows the partonic fractions for the analytical resummation without the changes to the colour evolution, for $Y$=5, $Q$=500$\,\mathrm{GeV}$ and $\sqrt{s}$=14$\,\mathrm{TeV}$. We see that it is mainly quark-quark interactions that dominate. There are two reasons for this, the first of which is that we are in the high $x$ regime ($x\sim1$) where the quark PDFs dominate over the gluon PDFs. The second reason is that gluons have a greater colour charge and therefore tend to radiate more, which results in less events involving gluons that contain an empty gap region. The plot is cut off at $Q_0$=1$\,\mathrm{GeV}$ as beyond that point it is mainly the non-perturbative behaviour which dominates and this is not implemented in the analytical calculation. The lower plot in Figure \ref{fig:fractionplot} shows the partonic fractions for the analytical resummation with the changes to the colour evolution, again for $Y$=5, $Q$=500$\,\mathrm{GeV}$. We see that there is a large difference here, and that now the processes involving gluons begin to dominate for low veto scales. It is this quasi-singlet behaviour which is generating the large partonic fraction for the $gg\rightarrow gg$ process.

\begin{figure}[!h] 
\begin{center}
\includegraphics[scale=0.68,angle=270]{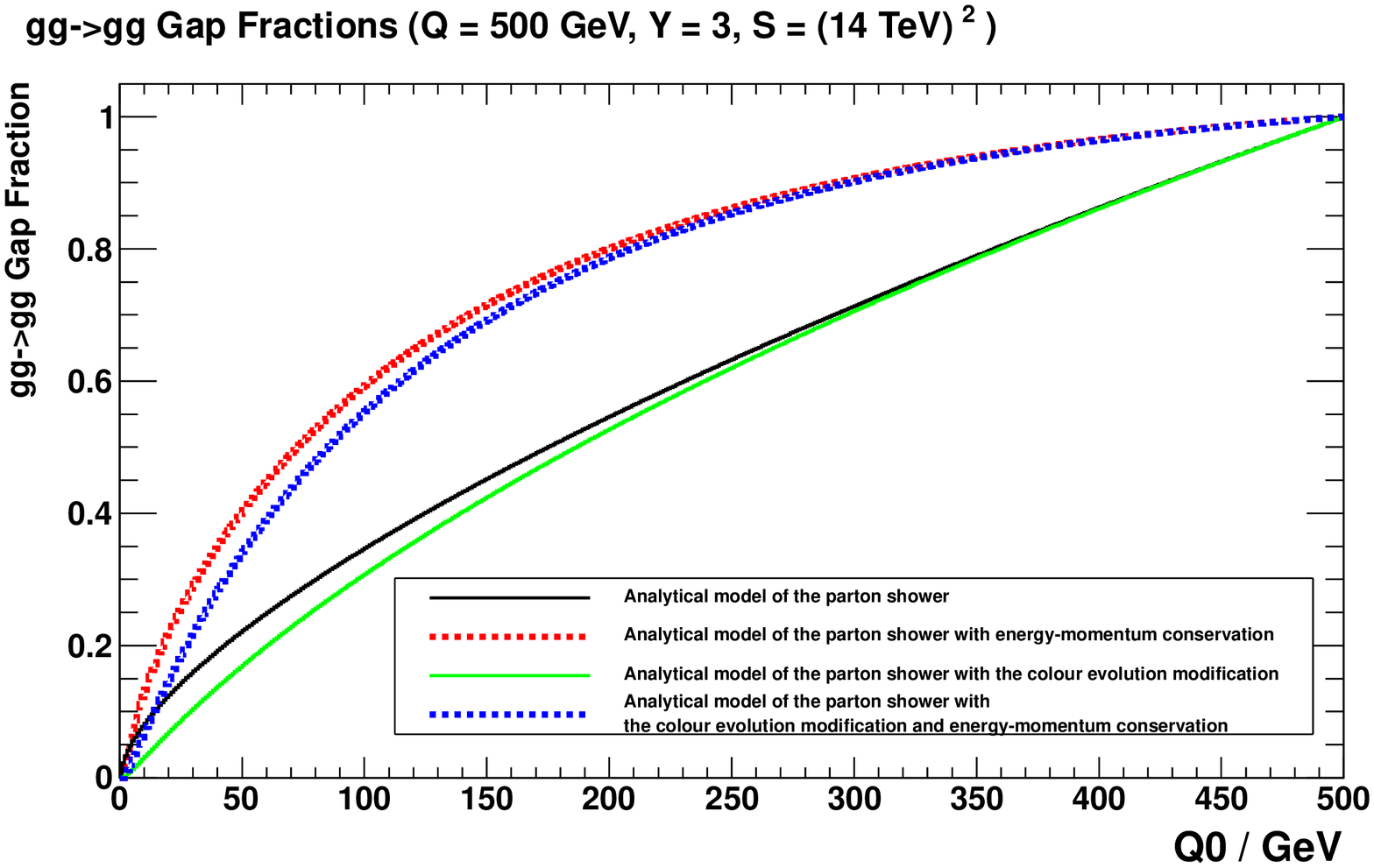}
\caption{Effects of partial energy-momentum conservation in the analytical calculation on the $gg\rightarrow gg$ gap fraction. The implementation of energy-conservation increases the gap fraction for low $Q_0$, as observed in the transition from black to red and green to blue respectively.}
\label{fig:consvsnocons}
\end{center}	
\end{figure}

In addition to the change in colour structure we also need to take into account effects of energy-momentum conservation. In the eikonal approximation one assumes that gluons which are emitted from the hard partons are infinitely soft and do not change the kinematics of the Born process. This correctly produces the leading logarithms we require for the resummation. In reality any emitted gluon carries a finite energy and momentum and this should be taken into account. This affects the PDFs, since more energy is needed in the hard scattering, and the hard partonic cross section, since recoil adjusts the kinematics. To estimate the effects of energy-momentum conservation we consider the case of a single emission of a gluon of energy $k_t$ and rapidity $y'$. This modifies the arguments of the PDFs to be

\begin{equation}
	\overline{x}_{1,2} = x_{1,2} + \frac{k_t e^{\pm (y'+\tfrac{\overline{y}}{2})} } {\sqrt{s}},
\end{equation}
where $x_{1,2}$ are the arguments of the PDFs without imposing energy-momentum conservation, $\sqrt{s}$ is the hadronic CMS energy and $\overline{y}$ is the sum of the rapidities of the two hard partons. We are no longer able to factorise out the $k_t$ and $y'$ integrals previously contained in the anomalous dimension from the PDFs. Instead of this we consider the simpler case where $k_t$=$Q_0$ and $y'$=$Y/2$, which will give an approximation of the maximum effect of a single emission. We then perform the calculation as before with the factorization, only now with the overestimate for the PDF arguments

\begin{equation}
	\overline{x}_{1,2} = x_{1,2} + \frac{Q_0 e^{\pm \tfrac{Y + \overline{y}}{2}}}{\sqrt{s}} .
\end{equation}
The effects of implementing partial energy-momentum conservation in the analytical calculation on the $gg\rightarrow gg$ gap fraction are shown in Figure \ref{fig:consvsnocons}. We define the gap fraction to be the ratio of events with a gap to the total number of events. We see that after implementing energy-momentum conservation there is a much larger gap fraction for lower $Q_0$. This is because higher energy emissions are suppressed by increasing $\overline{x}(Q_0)$ in the PDFs.  Note that while we have overestimated the effects due to one emission, we have neglected the effects due to multiple gluon emissions, and in some sense have underestimated the effects of energy-momentum conservation.

\section{Modifications to the parton shower}
\label{sec:psmodification}

Using the insight we have gained from the analytical calculation we now understand how the parton shower behaves relative to what we would have expected. The current parton shower generates an excess of wide-angle radiation 50\% of the time and no wide-angle radiation the other 50\% of the time. This leads to a larger number of events passing the veto on radiation than one would expect. To correct this behaviour we choose to modify the internal structure of the parton shower. 

Our aim in changing the parton shower is to allow each line of a gluon to radiate quasi-independently i.e. in a more similar way to the dipole shower formulation. The first modification is to change the assignment of partners after the hard scattering so that the partner with the largest angle is always chosen. When this is done the shower records the scale with respect to the furthest parton, $\tilde{q}_f$, and the scale with respect to the other parton, $\tilde{q}_n$. The shower then begins with half the colour factor, $\tfrac{1}{2} C_A$, as only one of the colour lines can radiate at the largest angle. The colour structure is set up such that the emitted gluons are only attached to the wide-angle colour line. Once the scale has been evolved down to $\tilde{q} < \tilde{q}_n$ then both lines are able to radiate and the colour factor is restored to $C_A$. The shower will attach the colour lines of the additional gluon with a 50-50 chance to one of the two lines of the parent.

\begin{figure}[!h] 
\begin{center}
\includegraphics[scale=1.0]{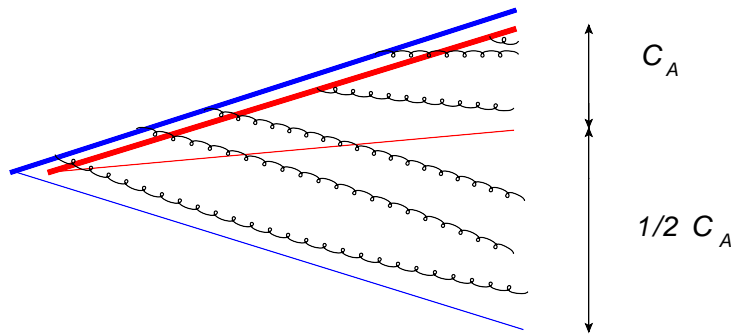}
\caption{The radiation pattern for the modified Herwig++ shower. The blue and red lines are separated by wide and small angles respectively. The bold lines are the colour lines of the gluon and the faint lines are the colour lines of the partners.}
\label{fig:patterns2}
\end{center}	
\end{figure}

The radiation pattern for the modified parton shower is shown in Figure \ref{fig:patterns2}. Note that the emitted gluons are connected to the correct lines, which results in a change to the hadronization behaviour.

\section{Results}
\label{sec:results}

For our investigation we choose to use the MSTW08lo PDF set \cite{Martin:2009iq} and use SISCONE \cite{Salam:2007xv} via FastJet \cite{Cacciari:2005hq} with cone radius $R$=0.4 and overlap parameter 0.5 for jet finding, as was done in the previous analysis. For the first two sets of results we turn off hadronization and multiple interactions, since these are not simulated in the analytical approach.

Now that the corrections to the parton shower have been made we will validate the approach by considering the $gg\rightarrow gg$ gap fraction, which is the most sensitive to the way radiation is treated in the parton shower. In the absense of energy-momentum corrections the gap fraction is simply the ratio of the resummed cross section to the Born cross section. There are three choices of variable that we can use in order to investigate the gap fraction, the hard scale $Q$, the rapidity gap size $Y$ and the veto scale $Q_0$. Changing $Q$ or $Y$ will modify the PDFs, the hard components and the colour suppression due to the large logarithms. In order to investigate the effects due to the change in the colour structure only, it is therefore better to investigate the gap fraction using $Q_0$, which changes only the large logarithms and, more weakly, the PDFs. Using $Q_0$ as a variable has previously been suggested for other processes \cite{Cox:2010ug}.

\begin{figure}[!h] 
\begin{center}
\includegraphics[scale=0.65,angle=270]{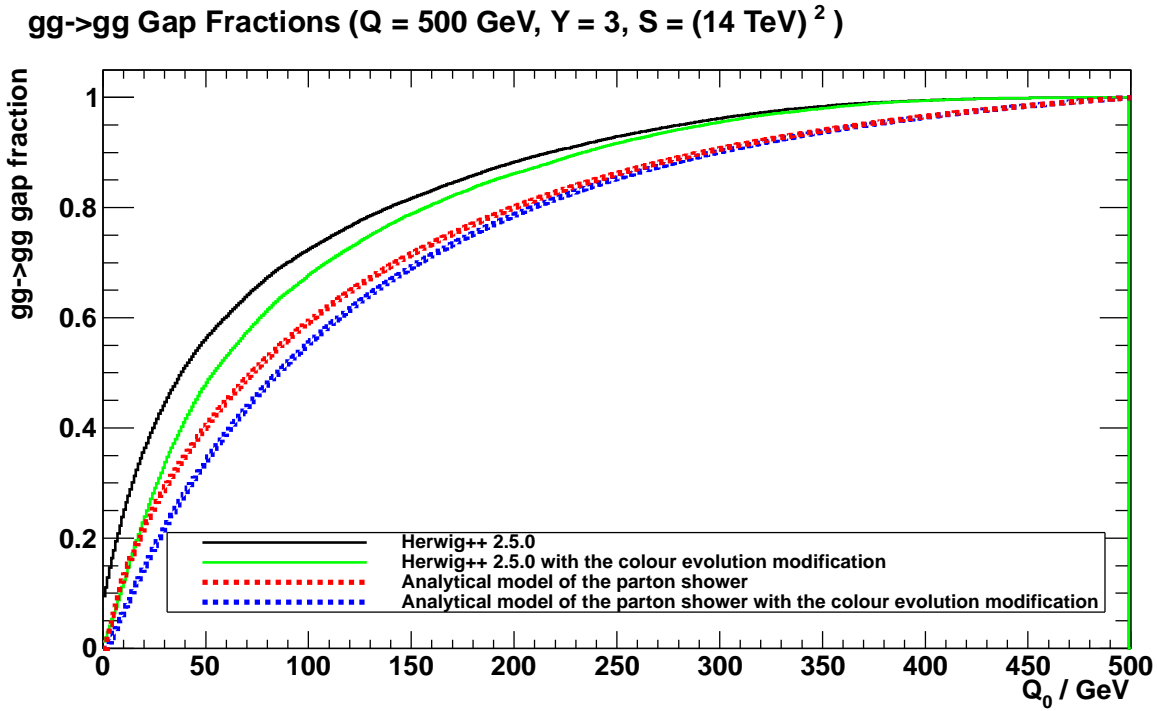}
\caption{The gap fraction as a function of $Q_0$. The red and black curves are the numerical results with and without the modifications to the colour evolution. The green and blue curves are the analytical results with and without the modifications to the colour evolution, both including the modifications to account for energy-momentum conservation described in Section 4.}
\label{fig:Comparison}
\end{center}	
\end{figure}

\begin{figure}[!h] 
\begin{center}
\includegraphics[scale=0.85]{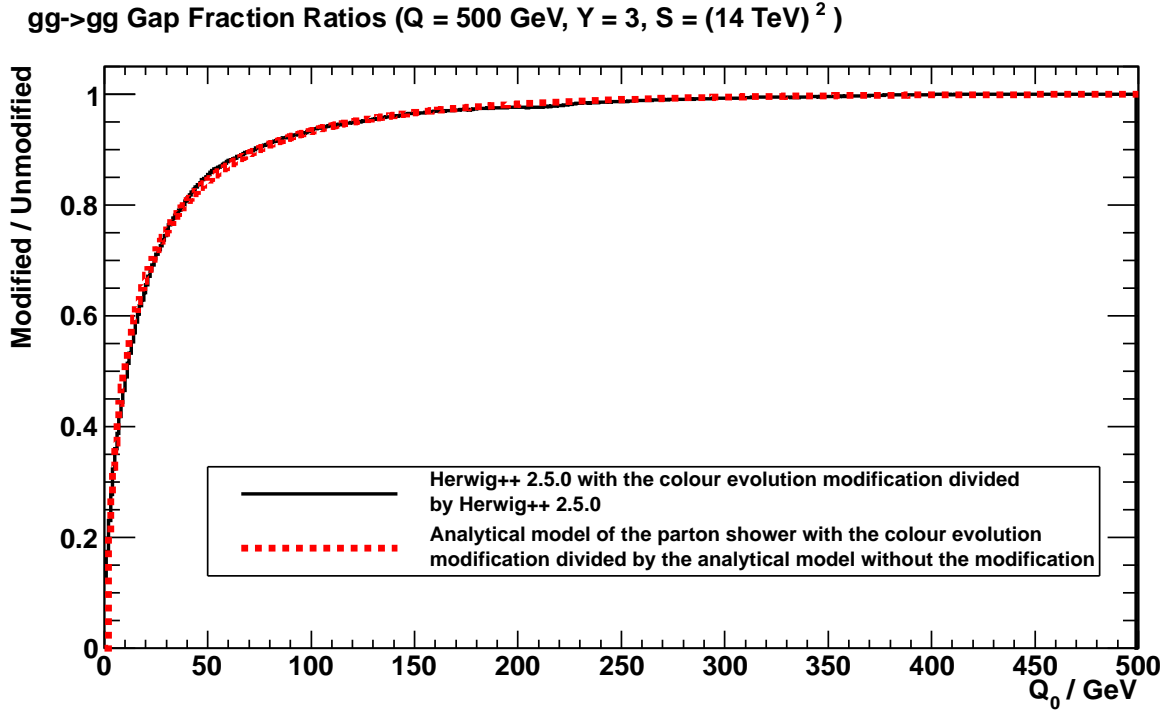}
\caption{The ratio of the colour evolution modified $gg\rightarrow gg$ gap fractions to those without modifications for Herwig++ 2.5.0 (black) and the analytical model of the parton shower (red).}
\label{fig:Comparison2}
\end{center}	
\end{figure}

\begin{figure}[!h] 
\begin{center}
\includegraphics[scale=0.85]{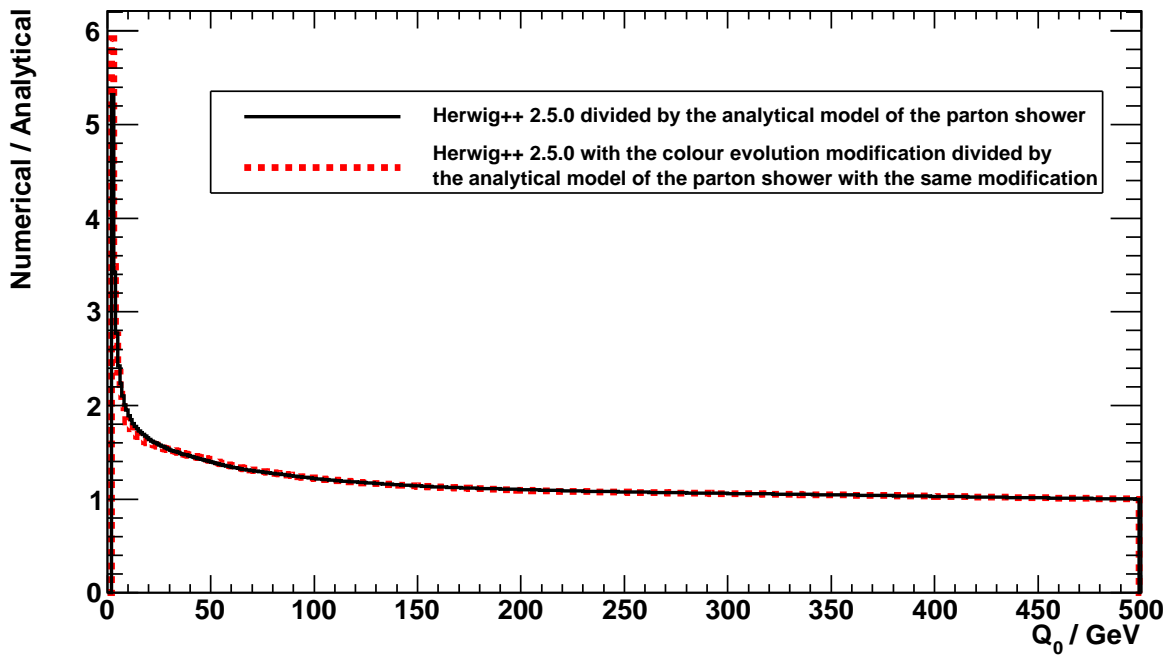}
\caption{The ratio of Herwig++ 2.5.0 to the analytical model of the parton shower with the colour evolution modification (black) and without (red).}
\label{fig:Comparison3}
\end{center}	
\end{figure}

Having decided upon the choice of variable, we now plot the gap fraction as a function of $Q_0$ in Figure \ref{fig:Comparison}. We choose the parameters $Y$=3, $Q$=500$\,\mathrm{GeV}$ as this enhances the logarithmic wide-angle suppressions and take the beam energy to be $\sqrt{s}$=14$\,\mathrm{TeV}$. It is clear that none of the curves seem to match and the analytical calculation generates more high energy radiation than Herwig++. Since we are interested in validating our change to the parton shower and not the gap fraction itself, we can remove any additional differences between the numerical and analytical approaches by forming ratios. Two different methods of dividing out are shown in Figures \ref{fig:Comparison2} and \ref{fig:Comparison3}. From the ratio plots we can conclude that, although the analytical curves do not match the numerical curves, our modification performs as expected. There are some minor disagreements in the low $Q_0$ region but since this region is dominated by non-perturbative effects we can no longer trust the analytical calculation, and the behaviour in the parton shower begins to be influenced by the infra-red cut-off.

\begin{figure}[!h] 
\begin{center}
\includegraphics[scale=0.85]{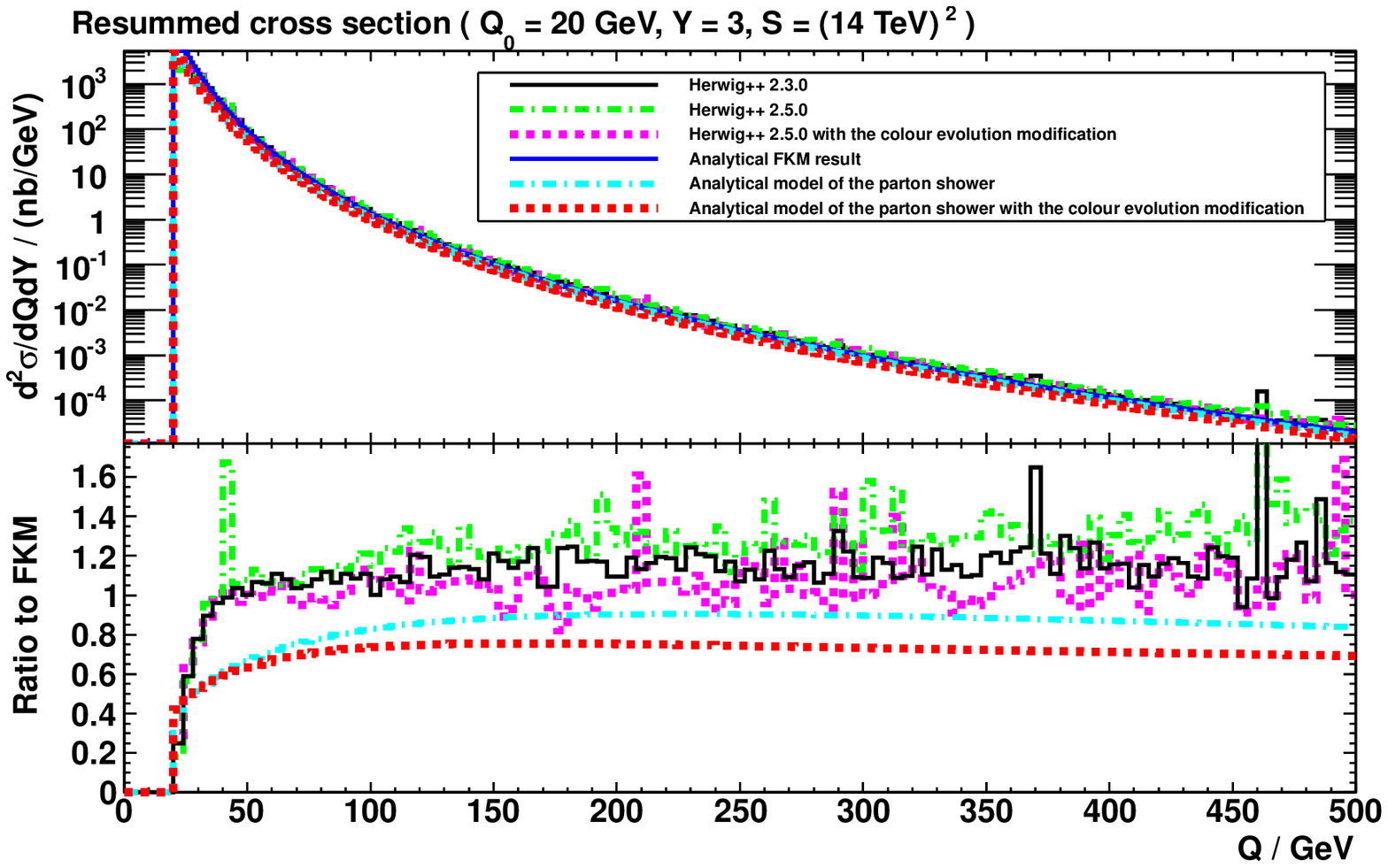}
\caption{The gap cross section for all possible QCD $2\rightarrow2$ processes as a function of the hard scale $Q$.}
\label{fig:QComparison}
\end{center}	
\end{figure}

Having convinced ourselves that the modifications have the correct physical behaviour, we now move onto the gap cross section as a function of $Q$, which was the original motivation for our present analysis. This is shown in Figure \ref{fig:QComparison}. The upper plot shows the resummed gap cross section, and the lower plot shows the ratio of the gap cross sections to the full colour structure result of Forshaw et al. \cite{Forshaw:2009fz}. The analytical predictions for the parton shower with and without our modification are clearly below that of the FKM result throughout the whole range of $Q$. Again we see a difference between the numerical and analytical results, with the numerical results always being greater than their analytical counterparts. Performing the same ratio analysis as above we have found that the modification behaves as expected.

\begin{figure}[!h] 
\begin{center}
\includegraphics[scale=0.65,angle=270]{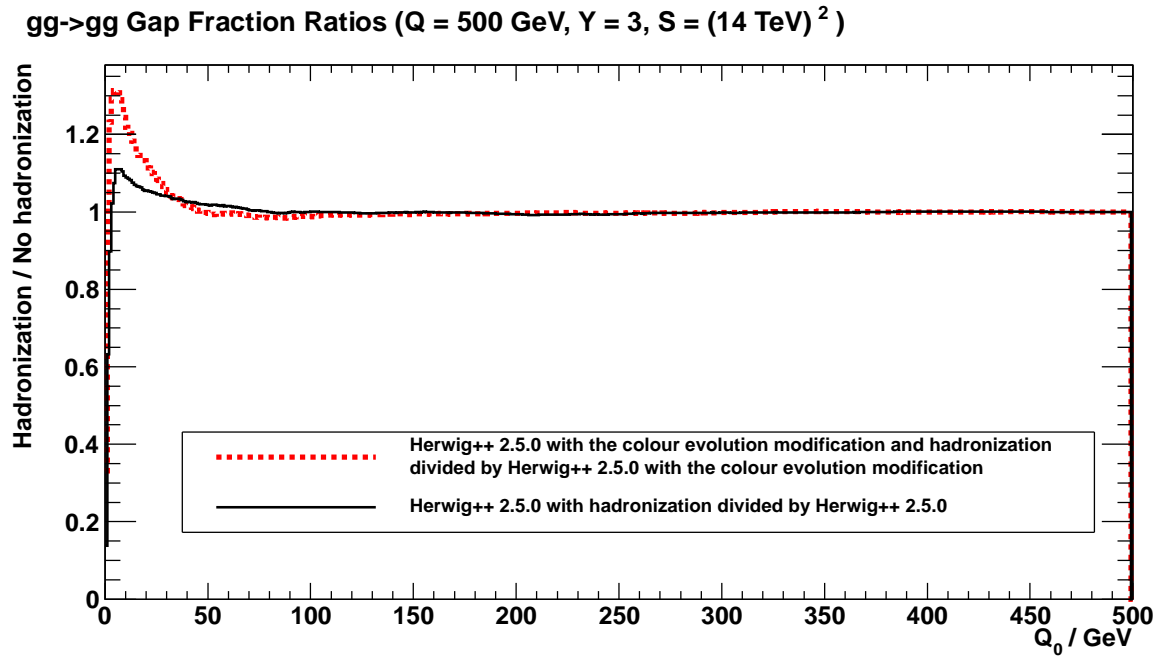}
\caption{The ratio of the $gg\rightarrow gg$ gap fraction generated by Herwig++ with hadronization to that generated by Herwig++ without hadronization.}
\label{fig:ComparisonH}
\end{center}	
\end{figure}

So far we have only considered the change in behaviour after running the parton shower. Since partons are not observed in the final state we should also include the effects of hadronization. These are shown in Figure \ref{fig:ComparisonH}. We see that including the hadronization effects change the gap fraction by a small amount relative to the parton shower level, with noticeable effects in the low $Q_0$ region. We do expect a difference because, comparing Figures \ref{fig:patterns} and \ref{fig:patterns2}, we see that the current algorithm has the effect of ``pulling'' the colour line from the small-angle scattering out into the gap region, which does not occur in the modified algorithm that we have presented.

\section{Conclusion}
\label{sec:conclusion}

Herwig++ is very successful at making a correspondence between hard processes and experimental observables. This is mostly due to the parton shower, which handles the evolution of a hard $2\rightarrow2$ scattering into experimentally observed high multiplicity final states. The underlying mechanisms of the parton shower are obtained from our understanding of the theoretical nature of QCD. As such, it is important that we investigate and correct any possible deficiencies in the parton shower approach. 

In this paper we have shown that the current parton shower of Herwig++ did not produce the right amount of wide-angle radiation for inclusive events. We interpreted this behaviour using an analytical model of the parton shower behaviour in gaps between jets events. This led to a modification of the internal structure of the parton shower. Although the changes we have made may have small effects in the current phase space the physical behaviour of the parton shower is now more in line with the analytical picture of radiation arising from QCD.

We would expect similar conclusions to apply to other parton shower algorithms (HERWIG and Pythia6), but more recent algorithms based on a dipole picture already include the effect we have considered.

Whether the parton shower is still successful with regards to predictions in gaps between jets events at the LHC has yet to be determined. Both the numerical and analytical approaches have deficencies with regards to each other. The current level of data analyses at the LHC \cite{ATLAS-CONF-2010-085} is not enough to distinguish between the two approaches. As more data arrive we will be able to see which of the two approaches best describes nature. 

From a theoretical point of view however, the parton shower is still lacking the inclusion of physical singlets responsible for a significant fraction of the cross section at large $Y$ and $Q$. These are not included since they involve evolution at the amplitude level, which is not possible as the current parton shower evolves the hard scattering at the cross section level. A full treatment of QCD in the parton shower will therefore require an amplitude level parton shower.

The code to implement our modificatins has been released in Herwig++ version 2.5.2, and may be switched on using the instructions in the Appendix.

\begin{flushleft}{\bf Acknowledgments}\end{flushleft} 
We thank the authors of Herwig++ for helpful comments, R. Duran Delgado and S.~Marzani for the code used in the analytical calculation, and J. Keates for the Herwig++ analysis code.

\bibliographystyle{utphys}
\bibliography{Jet_Vetoing_Herwig}

\appendix

\section{Switches}

To enable the new colour evolution model in Herwig++ 2.5.2, add the following lines to the ``..in'' file used with Herwig++

\begin{verbatim}
cd /Herwig/Shower
set Evolver:ColourEvolutionMethod 1
set PartnerFinder:PartnerMethod 1
set GtoGGSplitFn:SplittingColourMethod 1
\end{verbatim}

\end{document}